\documentclass{article}

\usepackage{arxiv}

\usepackage[utf8]{inputenc} 
\usepackage[T1]{fontenc}    
\usepackage{hyperref}       
\usepackage{url}            
\usepackage{booktabs}       
\usepackage{amsfonts}       
\usepackage{nicefrac}       
\usepackage{microtype}      
\usepackage{amsmath}
\usepackage{cleveref}       
\usepackage{lipsum}         
\usepackage{graphicx}
\usepackage{natbib}
\usepackage{doi}

\title{Too Few or Too Many? Sample Size Estimation for Differential Abundance Studies}

\date{15 June 2026}

\newif\ifuniqueAffiliation
\uniqueAffiliationtrue

\ifuniqueAffiliation
\author{ \href{https://orcid.org/0009-0007-2655-4094}{\includegraphics[scale=0.06]{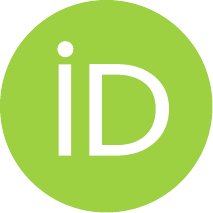}\hspace{1mm}Michael Agronah} \\
  Department of Mathematical and Statistical Sciences \\ 
  University of Alberta \\
  University Commons 5-140, T6G 2N8 \\
  Edmonton, Alberta, Canada \\
	\texttt{agronah@ualberta.ca}
	\And
	\href{https://orcid.org/0000-0002-2127-0443}{\includegraphics[scale=0.06]{orcid.pdf}\hspace{1mm}Benjamin M. Bolker} \\
  Departments of Mathematics \& Statistics and Biology \\
  McMaster University \\
  1280 Main Street West, L8S 4K1 \\
  Hamilton, Ontario, Canada}

\renewcommand{\shorttitle}{Sample size estimation}

\hypersetup{
pdftitle={Sample size estimation},
pdfsubject={q-bio.QM},
pdfauthor={Michael Agronah, Benjamin M. Bolker}
pdfkeywords = {Sample Size, Statistical Power, Differential Abundance, Microbiome}
}

\begin{document}
\maketitle

\begin{abstract}
  Determining an appropriate sample size for a study is a crucial step in planning scientific research. Appropriate sample size planning avoids both inadequate and inflated sample sizes.  
Inflated sample sizes wastes resources, time and effort of human subjects, and lives of experimental animals. Inadequate sample sizes, a much more common problem, wastes even more resources through the inability to detect biologically meaningful differences and encourages questionable research practices like $p$-hacking.  
Microbiome studies are particularly challenged by small sample sizes, particularly in studies of human subjects or expensive animal models. 
In practice, the statistical power of taxa within a differential abundance study is influenced by the effect size (typically quantified as fold change), mean abundance of individual taxa, and the number of samples. 
We present a novel approach for sample size calculation for differential abundance studies as a function of effect size, mean abundance and statistical power of taxa. 
Our method is implemented in the \texttt{power.nb} \textsf{R} package, available at \href{https://michaelagronah.com/power.nb/articles/stub.html}{https://michaelagronah.com/power.nb/articles/stub.html}. 
We applied our model for sample size calculation using estimates of mean abundance and fold change of taxa obtained from thirty real-world microbiome datasets. Our results showed that differential abundance microbiome studies require larger sample sizes than are currently prevalent in the literature to achieve adequate statistical power. Our framework will help researchers make informed decisions about appropriate sample sizes.
\end{abstract}


\section{Introduction}\label{sec1}
Choosing an appropriate sample size is a fundamental step in the design of 
scientific studies \citep{singh2014sampling}.
Careful sample size determination helps avoid problems associated with both 
over- and under-sampling. 
Excessively large sample sizes can lead to unnecessary expenditure, waste of 
time and resources, and the overuse of human participants or experimental 
animals \citep{button2013power}. Conversely, insufficient sample sizes reduce 
the ability to detect biologically meaningful differences \citep{button2013power}.

Microbiome studies are particularly prone to challenges arising from limited 
sample sizes \citep{kers2021power}. Small-sample studies typically exhibit low
statistical power, increasing the likelihood of failing to detect true biological 
signals. In addition, effect size estimates from such studies tend to be highly
variable, making it difficult to distinguish genuine biological differences 
from noise. A further consequence is the systematic overestimation of effect 
sizes among taxa deemed statistically significant, a phenomenon commonly 
referred to as the ``winner's curse'' \citep{button2013power}. 

Microbiome analyses can be broadly categorized into three classes. 
First, \emph{univariate summary approaches} reduce each sample to a scalar metric 
(e.g., alpha diversity) and test for differences between groups. 
Second, \emph{community-level methods} assess global compositional differences 
using techniques such as PERMANOVA or Dirichlet--multinomial models. 
Third, \emph{differential abundance analyses} aim to identify individual taxa 
that exhibit significant changes across conditions. Existing methodological 
developments for power and sample size estimation in microbiome studies have
largely focused on univariate summaries or community-level comparisons 
\citep{xia2018statistical, agronahnbolker}. 

Differential abundance microbiome studies present additional challenges for 
sample size determination compared with analyses based on overall community 
composition or univariate summary measures. In differential abundance settings,
statistical power varies across taxa because each taxon is characterized by its 
own effect size and mean abundance. Consequently, sample size requirements for
differential abundance analysis must account for taxon-specific mean abundance 
and fold change. Power analysis, widely used in scientific research for sample 
size determination \citep{cohen2013statistical,xia2018statistical}, can 
therefore be applied to guide appropriate sample size selection in differential
abundance microbiome studies. Statistical power in differential abundance studies 
is jointly influenced by sample size, taxon mean abundance, and effect size. 
Because microbiome data consist of counts, taxa with lower mean abundance exhibit 
higher relative variability, leading to reduced power for detecting differences. 
For example, detecting a 10\% change in a highly abundant taxon 
(e. g., from 1000 to 1100 reads) is substantially easier than detecting the 
same relative change in a rare taxon (e.g., from 10 to 11 reads), due to the 
higher variability associated with low counts.

Researchers often select sample sizes based on precedent from previous studies 
\citep{singh2014sampling}. This practice can be problematic, particularly in 
fields where studies are already underpowered. Reliance on historically small
sample sizes risks perpetuating low statistical power and producing unreliable 
or inconclusive results \citep{button2013power, ioannidis2005most}. 
Power analysis provides a principled framework for determining sample size 
\citep{cohen2013statistical, xia2018statistical}, but its application in 
microbiome differential abundance studies requires careful consideration of 
taxon-level variability in effect size and mean abundance.

To address these challenges, \cite{agronahnbolker} proposed a method for 
estimating statistical power for individual taxa as a function of mean 
abundance and effect size. A limitation of this framework is that it does not 
explicitly include sample size as a predictor of statistical power or model 
interactions between sample size, mean abundance, and effect size (fold change). 
Power is estimated as a function of mean abundance and fold change, conditional 
on a fixed sample size, and thus the framework does not support estimation of 
the sample size required to achieve a target power.  Building on this framework, 
the present study incorporates sample size as an additional determinant of statistical
power and develops a novel approach for estimating the sample size required to achieve 
a desired level of power for taxa with specified mean abundance and effect size. 
By explicitly modeling pairwise interractions between these factors, our method 
provides a practical tool to guide study design in differential abundance microbiome 
research and enables more informed decisions about appropriate sample sizes.

\section{Materials and Methods}
\subsection{Statistical Power Estimation}
Let $X$ denote an $m \times n$ Amplicon Sequence Variants (ASV) table with $n$ taxa and $m$ subjects and let $x_{1i}$ and $x_{2i}, \textrm{for}  \, i = 1, ..., n$, denote the log mean count for taxa $i$ (i.e., log of the average count across all subjects) and the log fold change (i.e. a measure of effect sizes) for taxa $i$, respectively.  
Define the event that the between-group difference of a given taxon is statistically significant at a particular significance level ($\alpha$) by the binary variable  $y$ (i.e., $0 \, \textrm{or} \, 1$), with 1 indicating  that the \( p \)-value for that taxon falls below a critical threshold (i.e., $p$-value $< \alpha$) and \(0\) otherwise. 

The power estimation method proposed by \cite{agronahnbolker}, is as follows: 

The variable $y$ follows a Bernoulli distribution with probability $p_i$. To estimate statistical power for various combinations of log mean count and log fold change, \cite{agronahnbolker} fitted a shape-constrained Generalized Additive Model (GAM) \citep{pya2015shape}. Defining the logistic function as
$$\textrm{Logist}(x) = \left(1 + \exp(-x)\right)^{-1},$$
the model formulation is given by 
\begin{equation}
\begin{cases}
y \sim \textrm{Bernoulli}(p_i) \\
p_i =   \textrm{Logist}(\eta) \\
\eta = \beta_0 + f(x_{1i}, x_{2i}), 
\end{cases} \label{eqn1}
\end{equation}
 
where the probability \( p_i \) represents the statistical power for taxon \( i \) and \( \beta_0 \) is the intercept.  The function \( f \) represents a two-dimensional smoothing surface, generated using a tensor product smooth of log mean abundance and log fold change.  For each combination of log mean count and log fold change, statistical power for each taxon can be predicted from the fitted GAM.

In this paper, we extend the model in eqn \ref{eqn1} to include sample size.  A challenge introduced by the inclusion of sample size is that sample size, log mean count and log fold change are related, and these relationship should be accounted. Additionally, fold change, sample size and mean count are positively correlated with statistical power. We therefore model pairwise interactions using individual monotonic two-dimensional smoothing surfaces, allowing statistical power to increase monotonically with sample size, effect size and mean count of taxa.  

Let $n_k$ for $k=1,2,...,N$ denote the sample size per group corresponding to the $k^{th}$ dataset. For a given dataset, we assume equal sample sizes across all groups. The value of $n_k$ is constant within each dataset but varies across datasets. The extended model is defined by

\begin{equation}
\begin{cases}
y &\sim \textrm{Bernoulli}(p_i) \\
p_i &=  \textrm{Logist}(\eta) \\
\eta &= \beta_0 + f_1({x}_{1i},{x}_{2i}) + f_2({x}_{1i},n_k) + f_3({x}_{2i},n_k), 
\end{cases} \label{eqn2}
\end{equation}
where the functions $f_1, f_2$ and $f_3$ are two-dimensional smoothing surfaces with basis generated by the tensor product smooth of pairs of $x_{1i},x_{2i}$ and $n_k$  \citep{pya2015shape}. 

We implemented the method describe in  eqn. \ref{eqn2} in the \texttt{power.nb} package. In practice, the model in eqn. \ref{eqn2} may occasionally fail to converge due to numerical instability. To address this, we fit both the full model and a reduced model that excludes interaction terms involving sample size. Specifically, in the reduced model, we define $\eta$ in eqn. \ref{eqn2} as $\eta = \beta_0 + f_1({x}_{1i},{x}_{2i}) + f_2(n_k)$. When both models converge, the final model is selected based on the Akaike Information Criterion (AIC).

\subsection{Sample Size Calculation}
For a given log mean abundance and log fold change, we estimate the sample size required to achieve a target power using a root-finding technique. 
Let $g(x_{1i}, x_{2i}, n_k)$ denote the estimated linear predictor from the fitted GAM. The predicted power is obtained via the logistic transformation

\begin{align}
\hat{p}_i = \textrm{Logist}\big(g(x_{1i}, x_{2i}, n_k)\big). 
\end{align}

Given ${x}_{1i}, {x}_{2i}$ and a target power $p^\ast$, we estimate the required sample size $n_k$ by finding the root of the equation:
\begin{align}
\textrm{Logist}\big(g(x_{1i}, x_{2i}, n_k)\big) - p^\ast = 0.
\end{align}

We performed the root-finding procedure using Brent's algorithm for unidimensional 
root finding implemented in the \texttt{uniroot} function in R \citep{R-base}.

\section{Data Simulation}  \label{realdata}
Statistical power for individual taxa cannot be reliably calculated analytically
using properties of count distributions due to the complexity of microbiome data. 
Consequently, as with most complex statistical models, power estimation for 
microbiome studies relies on simulation-based approaches that mimic realistic 
microbiome data \citep{arnold2011simulation}. We used the Mixture of Gaussians 
Simulation approach  \citep{agronahnbolker} to generate microbiome count data 
for power and sample size calculations. This method models the distributions of 
taxon-specific mean abundance and effect size (log-fold changes) using finite 
mixtures of Gaussian distributions. In particular, taxon mean abundances are 
modeled as a finite Gaussian mixture, while fold changes, expressed as functions
of mean abundance, are similarly modeled using a mixture of gaussian distributions. 
This framework enables simulation of realistic mean abundance and effect size 
combinations for estimation of taxon-specific statistical power.
The method is implemented in the \texttt{power.nb} R package 
\citep{agronahnbolkerpowernb}. To ensure that simulated data reflect realistic
microbiome data, we estimated the parameters of the mixture model from thirty 
real microbiome datasets analyzed in \cite{nearing2022microbiome}. 
These datasets were sampled across a diverse range of host organisms, 
environments, and experimental conditions, including human clinical samples, 
animal-associated microbiomes, and environmental communities such as soil 
and water. The Supplementary Material presents the distributions of mean 
abundances, fold changes, and the number of control and treatment sample sizes 
for each dataset.

\subsection{Pre-filtering and Simulation process}

Following standard procedures in differential abundance microbiome studies, 
we filtered rare taxa, retaining only those with an abundance of five or more 
reads in at least two samples \citep{xia2018statistical, love2014moderated}. 
We estimated thirty sets of model parameters from fitting the Mixture of 
Gaussians simulation model to each dataset.  For each set of parameter estimate,
we simulated mean abundances, fold changes and corresponding microbiome count 
datasets using 1,000 taxa, across two groups (control and treatment).  
Simulations were performed at seven sample sizes per group: 10, 25, 40, 55,
70, 85 and 100. For each sample size, we generated five replicate datasets.

To assess whether a taxon differs significantly between groups, we fitted a 
negative binomial model to each simulated count data using the \texttt{DESeq2} 
R package and computed $p$-values associated to each taxon. 
The Benjamini–Hochberg procedure was used to control the false discovery 
rate (FDR). In order to predict statistical power across combinations of mean 
abundance, fold change and sample size per group, we combined all simulated 
datasets into a large single dataset containing all simulated mean abundances 
and fold changes, sample sizes per group used for the simulations, and the 
binary indicator denoting whether each taxon differed significantly between 
groups. The GAM described in eqn. \ref{eqn2} was then fitted to predict 
statistical power as a function of mean abundance, fold change and sample size 
per group.

\section{Results and Discussion} 

\begin{figure*}%
\centering
\begin{minipage}{0.45\textwidth}
    \centering
    \includegraphics[width=\textwidth]{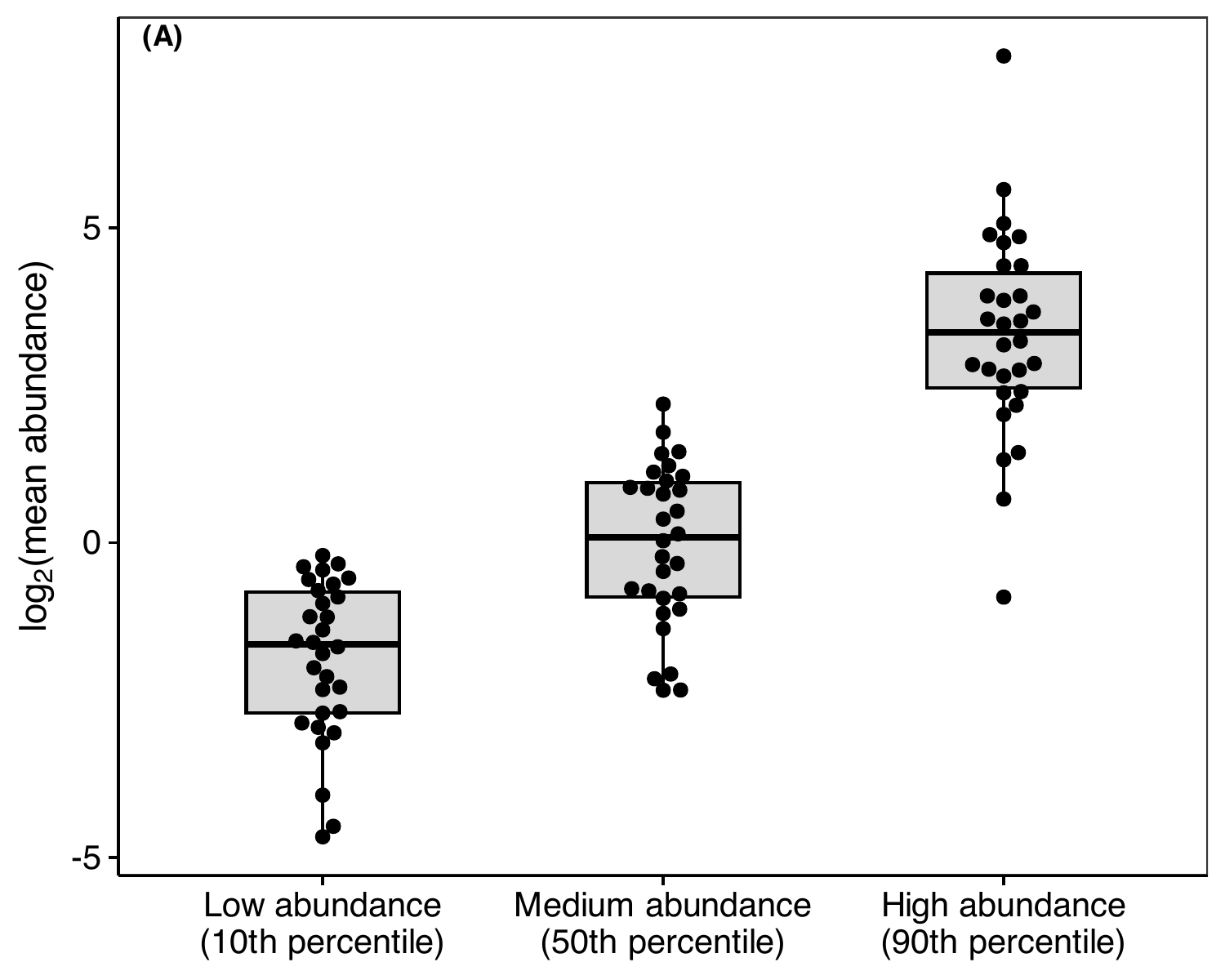} 
\end{minipage} \label{med_botha}
\begin{minipage}{0.45\textwidth}
    \centering
    \includegraphics[width=\textwidth]{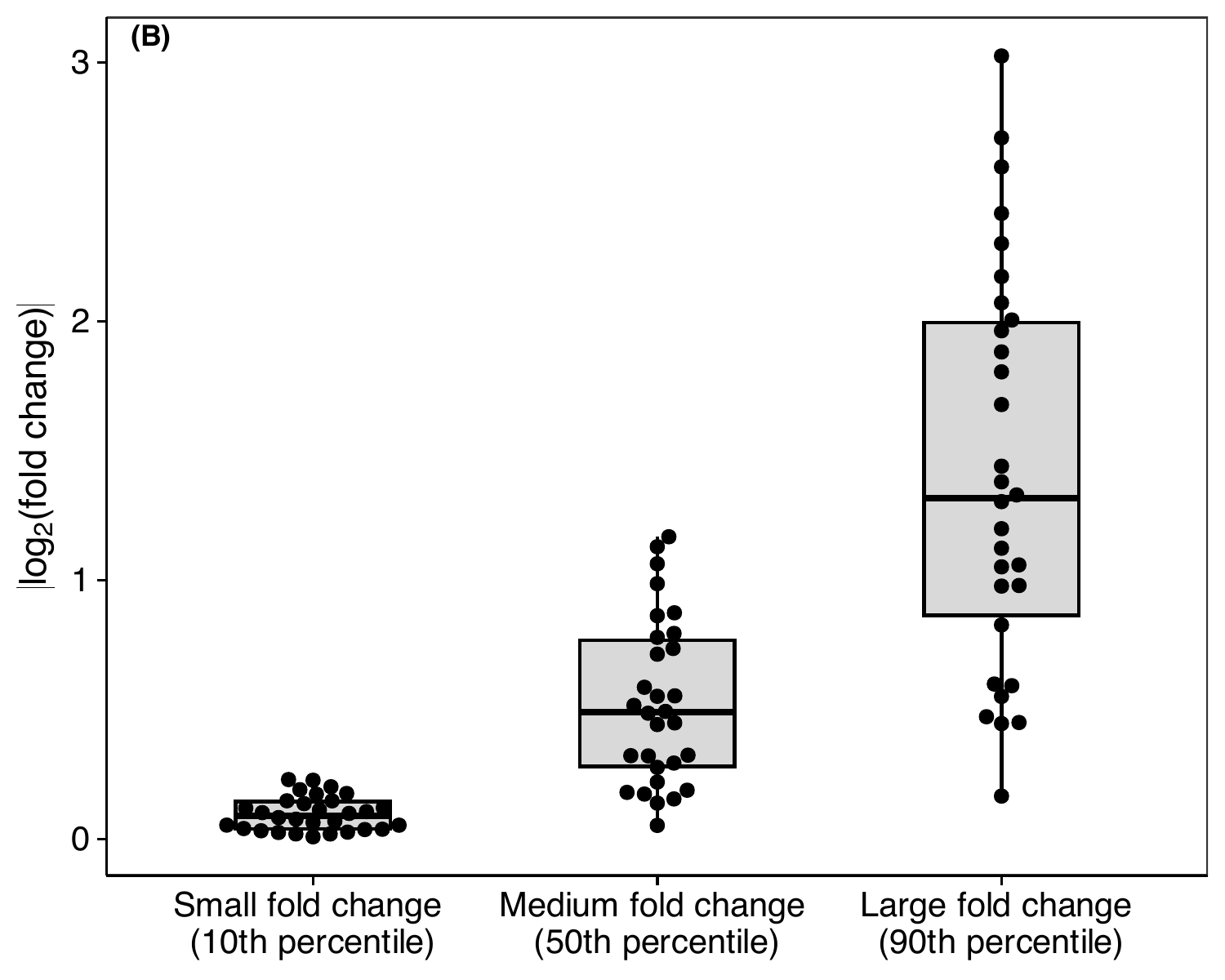} 
\end{minipage}
\caption{
Summary of mean abundance and fold change across the 30 microbiome datasets 
included in this study.
\textbf{(A)} Distributions of the 10th, 50th, and 90th percentiles of log$_2$ 
mean abundance computed within each dataset. The median values of these 
distributions were $-1.62$, $0.09$, and $3.34$, respectively, and were used to 
define low-, medium-, and high-abundance taxa in subsequent analyses.
\textbf{(B)} Distributions of the 10th, 50th, and 90th percentiles of the 
absolute log$_2$ fold changes estimated using DESeq2 within each dataset. 
The median values of these distributions were $0.09$, $0.49$, and $1.32$, 
respectively, and were used to define small-, medium-, and large-effect sizes. 
Points represent individual datasets and boxplots summarize variation across datasets.
}
\label{med_both}
\end{figure*}

Figure~\ref{med_both} shows the distributions of the 10th, 50th, and 90th 
percentiles of log$_2$ mean abundance and the magnitude of log$_2$ fold changes 
estimated using \texttt{DESeq2} across the 30 datasets. Representative levels of mean 
abundance and fold change  were defined as the medians of these percentile 
distributions across the datasets.
Low, medium, and high mean abundance correspond to the median 10th, 50th, 
and 90th percentiles of log$_2$ mean abundance, while small, medium, and 
large fold change correspond to the median 10th, 50th, and 90th percentiles of 
the absolute log$_2$ fold changes. The resulting values were $(-1.62,0.09,3.34)$ 
for mean abundance and $(0.09,0.49,1.32)$ for fold change. These values were 
subsequently used to evaluate the effects of mean abundance and fold change on 
statistical power and  sample size requirements.

\begin{figure*}%
 \centering
 \hspace*{0cm} 
    \includegraphics[scale=0.35]{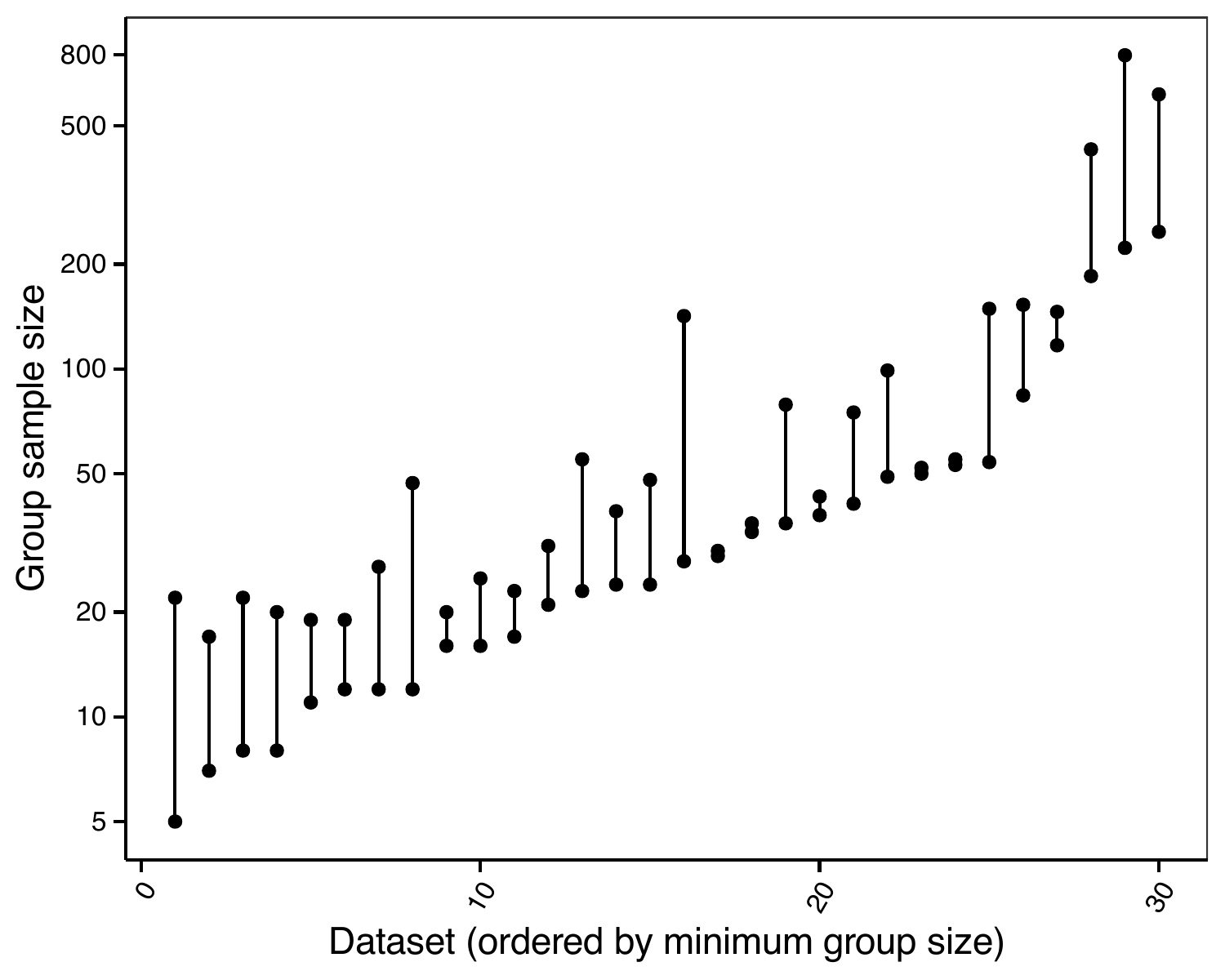} 
\caption{Observed group-size ranges across the 30 microbiome datasets analyzed. 
For each dataset, the lower and upper points denote the minimum and maximum group
sizes, respectively, and the connecting line indicates the range of group sizes 
within the dataset. Datasets are ordered by increasing minimum group size.}
\label{sample_ss_pplt}
\end{figure*}

Figure~\ref{sample_ss_pplt} shows the group sample sizes across 
the 30 datasets. For each dataset, the lower and upper points correspond to the
minimum and maximum group sizes, respectively, and the connecting line represents
the range of observed group sizes.
Datasets are ordered by increasing minimum group size.
Representative sample-size levels were defined using the 10th, 
50th, and 90th percentiles of the distribution of the largest group sizes 
across datasets. The resulting values ($20$, $45$, and $181$) were used as 
representative small, medium, and large sample-size settings, respectively, 
in subsequent analyses.

\subsection{Power estimates}

\begin{figure*}%
 \centering
 \hspace*{0cm} 
    \includegraphics[scale=0.6]{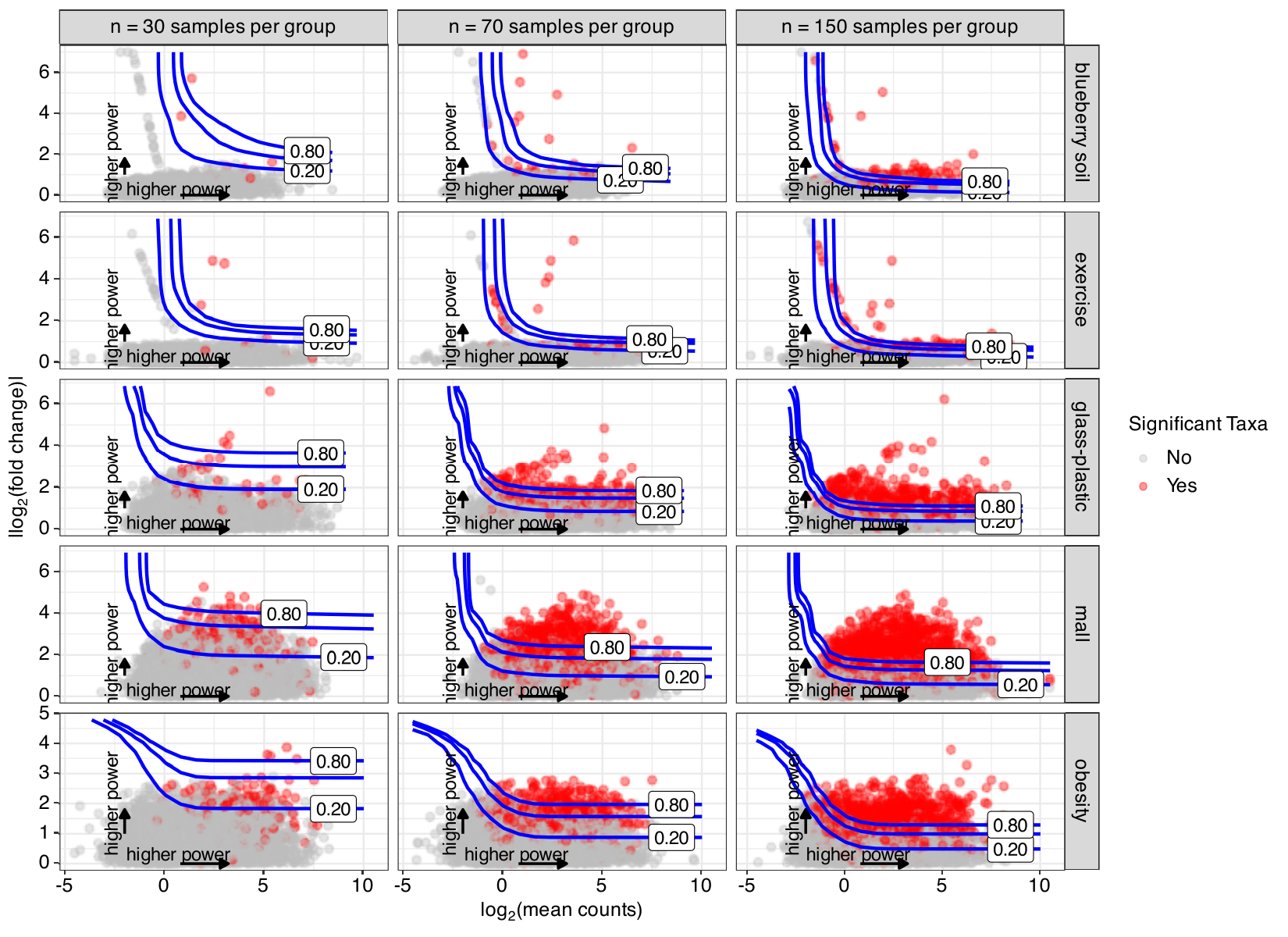} 
\caption{Relationship between statistical power for different combinations for mean count and fold change across sample size per group of $30, 70$ and $150$.}
\label{poww_plt}
\end{figure*}

Figure \ref{poww_plt} illustrates how statistical power varies as a function of 
log mean abundance and fold change across sample sizes for 5 selected datasets. 
The corresponding plots for the remaining datasets are provided in the 
Supplementary Materials.  
The red points indicate taxa that were identified as showing significant 
differences between groups after false-discovery-rate correction, while the gray 
points are taxa that did not pass the significance threshold.
We used the default false-discovery-rate threshold of 0.1 in the DESeq2 package,
which applies the Benjamini-Hochberg correction procedure.

The  contour curves show statistical power for various combinations of mean 
abundance and fold change. 

Across all five datasets, taxa with higher mean abundance or larger effect sizes
are more likely to be detected as significant. Rare taxa may still be detected 
when fold changes are sufficiently large, whereas highly abundant taxa can achieve 
significance even for modest fold changes. For each dataset, increasing sample 
size allows smaller fold changes to be detected at the same power level. 
Thus, larger sample sizes improve the ability to detect weaker differential 
abundance signals.
These results align with theoretical expectations that statistical power 
increases with effect size, mean abundance, and sample size 
\citep{agronahnbolker,xia2018statistical}.

\begin{figure*}%
  \centering
 \hspace*{0cm} 
    \includegraphics[scale=0.45]{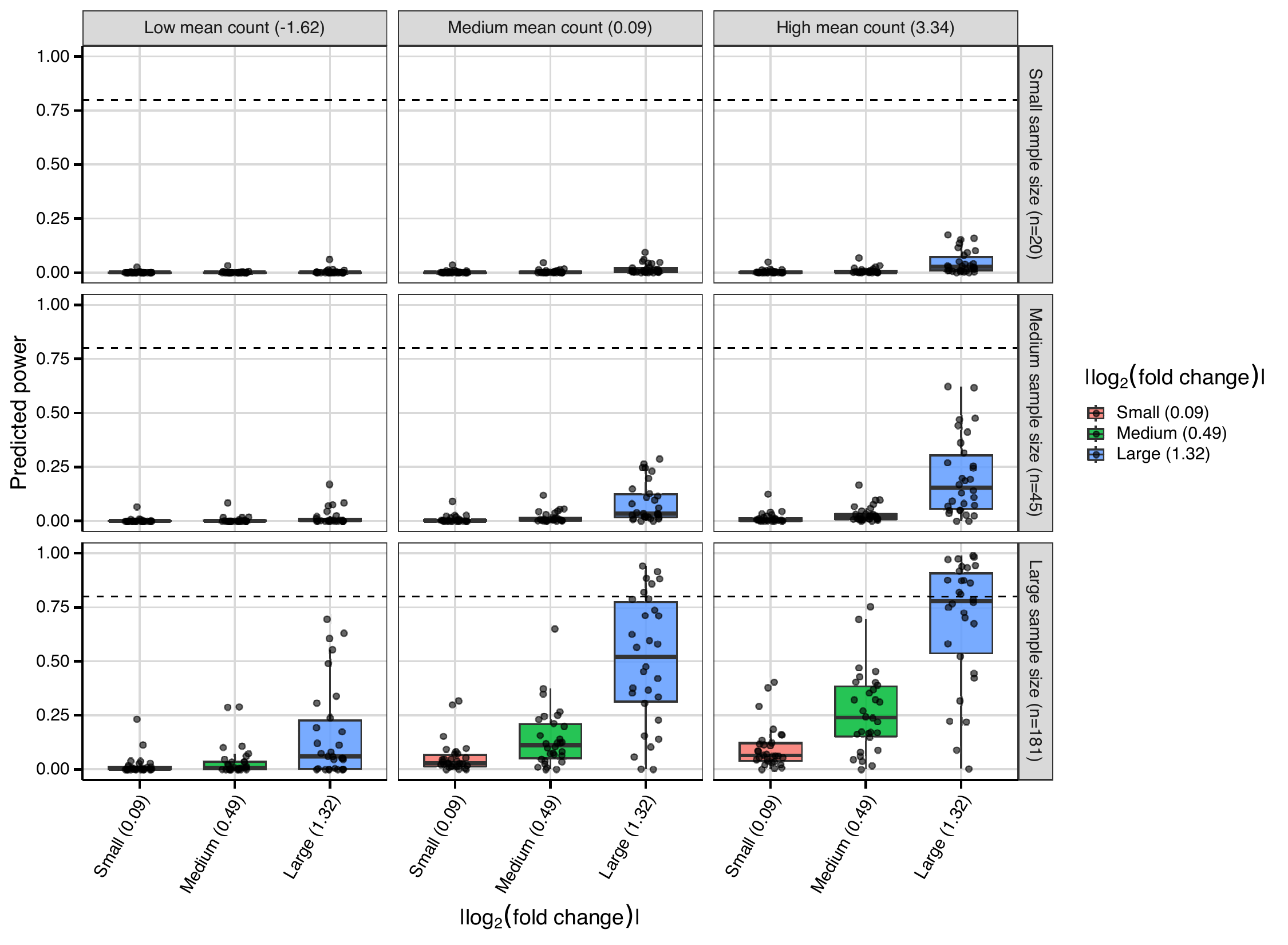} 
\caption{Predicted power as a function of mean abundance, effect size, 
and sample size. Predicted power is shown for combinations of 
$\log_2(\textrm{mean count})$, $|\log_2(\textrm{fold change})|$, 
and sample size across the 30 microbiome datasets. 
For $\log_2(\textrm{mean count})$, the low, medium, and high mean abundance levels 
($-1.62$, $0.09$, and $3.34$, respectively) correspond to the medians of the 
dataset-specific 10th, 50th, and 90th percentiles. Similarly, the small, medium,
and large effect-size levels for $|\log_2(\textrm{fold change})|$ 
($0.09$, $0.49$, and $1.32$, respectively) correspond to the medians of the 
dataset-specific 10th, 50th, and 90th percentiles. The small, medium, and large 
sample-size levels ($20$, $45$, and $181$ samples per group, respectively) 
correspond to the 10th, 50th, and 90th percentiles of the distribution of the 
larger of the treatment and control group sizes across the 30 datasets. 
Boxplots summarize predicted power across datasets, while points represent 
individual dataset predictions. Panels are arranged by mean abundance (columns) 
and sample size (rows). }
\label{fig:power_plot}
\end{figure*}

Figure~\ref{fig:power_plot} shows predicted statistical power as a function of 
mean abundance, fold change and sample size across the thirty microbiome datasets. 
Panels are arranged by increasing mean abundance (columns) and sample size (rows). 
Boxplots summarize variability across datasets, with points representing individual datasets. 
In practice, a statistical power of 80\% or higher is commonly regarded as sufficient power 
\citep{cohen2013statistical,descoteaux2007statistical}. 
For small to medium sample sizes per group ($n=20, 45$), predicted statistical 
power remains well below the conventional 80\% threshold for all datasets, even
at medium to high combinations of fold change and mean abundance. 
Even for highly abundant taxa, power falls short of commonly accepted levels,
indicating that differential abundance studies with limited sample sizes are
generally underpowered. 
At medium sample sizes $(n=45)$, power increases modestly for taxa with 
high mean abundance $(\log_2(\textrm{mean count}) = 3.34)$ and large effect size
$(|\log_2(\textrm{fold change})| = 1.32)$. However, power still remains well below
the 80\% power level. This suggests that sample sizes typical of many microbiome
studies are insufficient to reliably detect differential abundance, especially
for low-abundance taxa. Substantial improvements in power were observed only at
the large sample size ($n = 181$). 
For  $n = 181$, taxa with both medium-to-high abundance and large fold changes 
frequently achieved or exceeded the 80\% target power threshold. In contrast, 
low-abundance taxa continued to exhibit low power even at large sample sizes for 
most datasets.

\subsection{Sample Size Estimation} 

\begin{figure*}%
  \centering
 \hspace*{-1cm} 
    \includegraphics[scale=0.45]{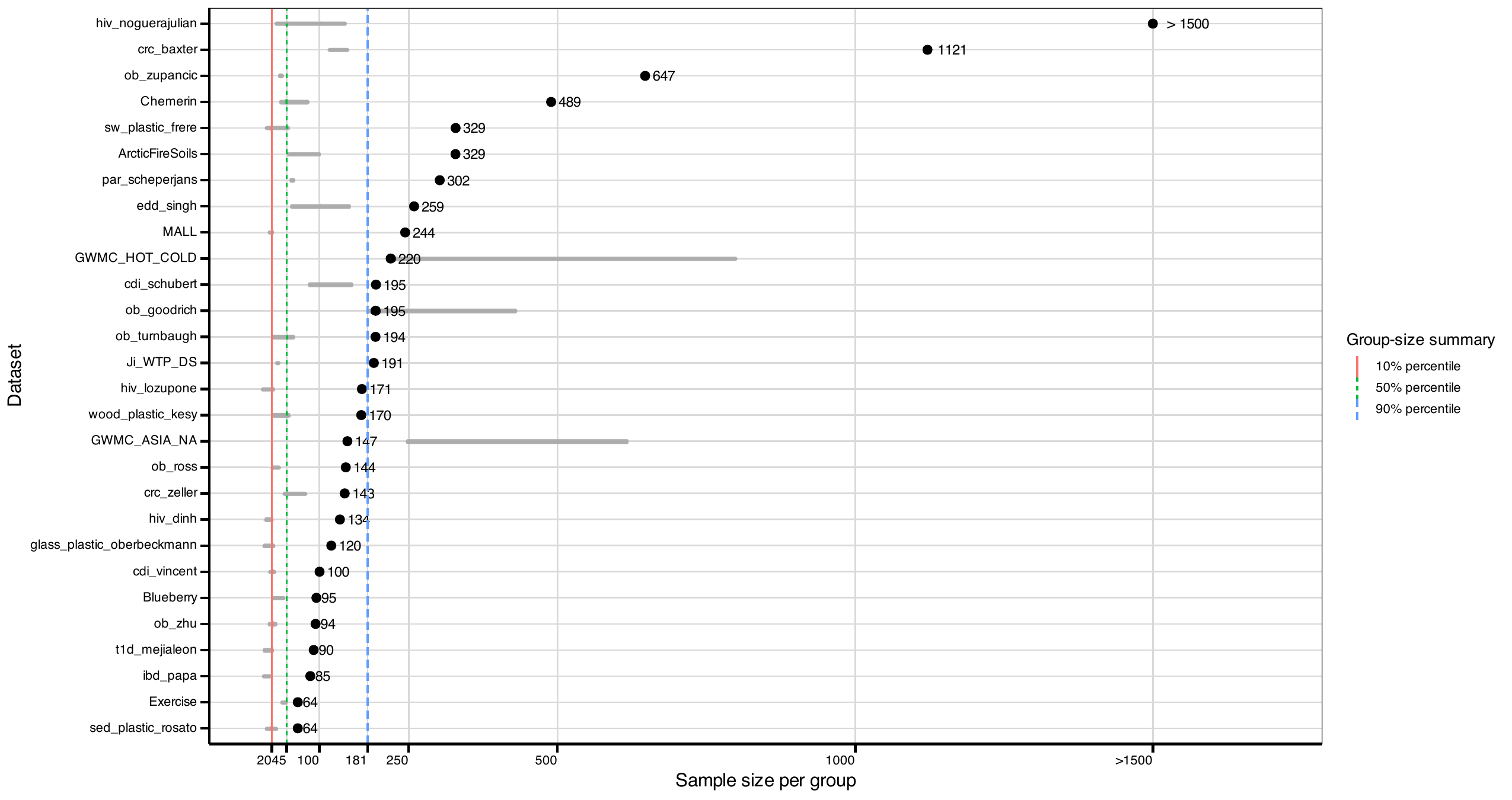} 
\caption{Estimated sample size required to reach 80\% target power
for taxa with large fold change 
$(|\log_2(\textrm{fold change})| = 1.32)$ and high mean abundance 
$(\log_2(\textrm{mean count}) = 3.34)$.
Grey horizontal bars show the observed range of group sizes within each dataset; 
black points show the estimated sample size per group required to achieve the 
target power. Vertical dashed lines indicate the minimum, median, and maximum 
observed largest group sizes across datasets.
Two datasets were excluded because no finite sample-size estimate was obtained 
within the prespecified search interval of 10 to 5000 sample sizes per group. 
For these datasets, the predicted power at the upper bound of the interval was 
only 0.18 and 0.72, indicating that substantially larger sample sizes would be 
required to achieve the target power.}
\label{fig:sample_size}
\end{figure*}

Figure~\ref{fig:sample_size} shows the estimated sample size per group required 
to achieve 80\% target power for taxa with large fold change 
$(|\log_2(\textrm{fold change})| = 1.32)$ and high mean abundance 
$(\log_2(\textrm{mean count}) = 3.34)$. 
The vertical reference lines indicate the 10th, 
50th, and 90th percentiles of the largest observed group sizes 
(between the control and treatment groups) across the thirty 
microbiome datasets. 
The grey horizontal bars show the observed range of group sizes within each 
dataset.

Many observed group-size ranges remain well below the estimated sample 
sizes required to achieve 80\% power.
The estimated sample sizes required to achieve the 80\% target power 
substantially exceeded the group sizes 
typically observed in microbiome studies. In particular, the majority of 
datasets required sample sizes well above the medium 
group size ($n = 45$), while several exceeded even the large  
group size ($n = 181$). Many datasets required 
more than 100 samples per group, several exceeded 200 samples 
per group, and two datasets surpassed the upper reporting 
threshold of 1500 samples per group.

These estimates correspond to taxa with high abundance and large fold changes, 
which are generally easier to detect in differential 
abundance analyses. Consequently, substantially larger sample 
sizes would be expected for taxa with lower abundance or smaller 
fold change. These findings therefore suggest that many existing 
microbiome differential abundance studies may be considerably 
underpowered, even for detecting relatively strong signals.

\section{Conclusion} 
We present a novel method for sample size calculation in differential abundance microbiome studies. 
While sample size is generally influenced by statistical power and effect sizes, in differential abundance studies, power also depends on mean abundance. 
Since statistical power is a function of both mean abundance and effect sizes of individual taxa, our approach models sample size as a function of these factors. 
Using our method, researchers can quantify the range of sample sizes required to detect taxa with various effect sizes and mean abundances at a target statistical power.  

To estimate statistical power, we apply the mixture of gaussians simulation method presented in \cite{agronahnbolker}. The method provides a flexible framework for modeling the distribution of mean abundance and effect size of taxa, which is essential for estimating taxon-specific statistical power. 
We fitted the mixture of gaussians simulation method to thirty real microbiome datasets to derive realistic parameter estimates for simulating microbiome datasets.

Our sample size estimation results show that larger sample sizes may be required than those commonly used in microbiome studies to achieve high statistical power to detect taxa with given effect sizes and mean abundances. This finding highlights the importance of conducting sample size calculations before a study begins, as relying on previous or pilot studies may be misleading especially if those studies already had low power.
Without adequate sample sizes, microbiome studies may fail to detect 
biologically meaningful differential abundance patterns, thereby 
contributing to low reproducibility and inconsistent findings 
across studies.

\subsection{Data and Code Availability}

All simulation code and scripts required to reproduce the analyses, figures, and results presented in this manuscript are publicly available at:

\href{https://github.com/magronah/Too-Few-or-Too-Many-Sample-Size-Estimation-for-Differential-Abundance-Studies/tree/main}{https://github.com/magronah/Too-Few-or-Too-Many-Sample-Size-Estimation-for-Differential-Abundance-Studies/tree/main}.

The proposed methodology is implemented in the \texttt{power.nb} R package, available at:

\href{https://michaelagronah.com/power.nb/articles/stub.html}{https://michaelagronah.com/power.nb/articles/stub.html}

\bibliographystyle{apalike}
\bibliography{reference}

\end{document}